\documentclass{article}

\usepackage[english]{babel}

\usepackage[letterpaper,top=2cm,bottom=2cm,left=3cm,right=3cm,marginparwidth=1.75cm]{geometry}

\usepackage{cite}

\usepackage{amsmath}
\usepackage{graphicx}
\usepackage[normalem]{ulem}
\usepackage[colorlinks=true, allcolors=blue]{hyperref}
\usepackage{amsmath,amssymb,epsfig,amsfonts,mathrsfs}
\DeclareSymbolFontAlphabet{\mathrsfs}{rsfs}

\newcommand{\beq}{\begin{equation}}
\newcommand{\eeq}{\end{equation}}
\newcommand{\ba}{\begin{aligned}}
\newcommand{\ea}{\end{aligned}}
\newcommand\ii{\mathrm{i}}

\newcommand\ve{\varepsilon}

\newcommand\BR{\mathbb{R}}

\newcommand\BC{\mathbb{C}}
\newcommand\BZ{\mathbb{Z}}

\newcommand\CalP{\mathcal{P}}

\newcommand\CalH{\mathcal{H}}

\newcommand\CalN{\mathcal{N}}

\newcommand\CalM{\mathcal{M}}

\newcommand\CalX{\mathcal{X}}
\newcommand\CalY{\mathcal{Y}}
\newcommand\CalZ{\mathcal{Z}}

\title{Analytic continuation and supersymmetry}
\author{Nikita Nekrasov}

\begin{document}
\maketitle

\begin{center}
Simons Center for Geometry and Physics, \\
Stony Brook University, Stony Brook, NY 11794-3636, USA
\end{center}
\begin{abstract}
 \noindent
 
Real-valued parameters of quantum field theory, such as Planck constant $\hbar$, the coupling constants $g^{-2}, {\theta}, \ldots$, 
 temperature ${\beta}^{-1}$ and  spacetime metric ${\bf g}_{\mu\nu}$, chemical potentials or background gauge fields ${\bf A}_\mu^a$, can be made complex. In perturbative string theory, the worldsheet metric is sometime Lorentzian, somewhere Euclidean, and likely complex in between.  Less often, integral, or quantized, parameters, such as the number $n$ of flavors or species (as in replica method), the number of colors $N$ (as in matrix models, master field), spacetime dimension $D$ (dimensional regularization), or angular momentum $l$ (Regge poles) are continued to complex values.  We observe that many, if not all of these analytic continuations can be realized
 with conventional supersymmetric quantum field (or M-) theory.  The supersymmetry is softly broken near a defect supporting the theory being analytically continued in some of its parameters. 
 
 \vskip 0.5cm
 
 Somewhat abridged talk at Strings-Math'2022 in Warsaw. 
 \vskip 1cm
 
\end{abstract}

%\tableofcontents

\section{Introduction}

 It is interesting to study analytic continuations. 
It is especially interesting to study analytic continuations of quantum field theory. Usually the real-valued parameters, such as Planck constant $\hbar$, the coupling constants $g^{-2}, {\theta}, \ldots$, 
 temperature ${\beta}^{-1}$ and  spacetime metric ${\bf g}_{\mu\nu}$, chemical potentials or background gauge fields ${\bf A}_\mu^a$, can be made complex. 
 For example, to a quantum mechanical system with discrete spectrum Hamiltonian $\hat H$ we associate
 the thermal partition function
 \beq
 {\CalZ}({\beta}) = {\rm Tr} \, e^{-{\beta}{\hat H}} = \oint_{C}\, dz e^{-{\beta}z}\, {\rm Tr} G_{\hat H}(z)
 \label{eq:thermaltr}
 \eeq
where $G_{\hat H}(z)$ is the resolvent \cite{Takhtajan:2020hwl}, and $C$ is the contour circling around the spectrum ${\sigma}({\hat H})$. Using path integral 
\beq
{\CalZ}({\beta}) = \int_{L{\CalP}} \, Dp(t) Dq(t) \, e^{{\ii} \oint p dq - \oint H\left( p(t), q(t)\right) d{\beta}}\ ,
\label{eq:trbeta}
\eeq
we represent the partition function as an integral over the space $L {\CalP}$ of loops on a phase space, with the measure defined
using a complex-valued one form $d{\beta} = {\beta}^{\prime}(t) dt$ on $S^1$.  If we eliminate momentum by its equations of motion, e.g. for $H$ which is quadratic in $p$, the saddle point dominating \eqref{eq:trbeta}, does not, typically, fit in ${\CalP}$, since 
\[ p^{\rm saddle}_{i}(t) = \frac{1}{{\ii}{\beta}^{\prime}(t)} \, g_{ij} \frac{dq^{j}}{dt}   \, , \qquad g^{ij} = \frac{{\partial}^{2} H}{\partial p_{i} \partial p_j} \]
is complex unless ${\beta}^{\prime}(t) \in {\ii}{\BR}$. Once $p(t)$ can become complex, so can $q(t)$. One is led to the idea
that path integral contour $L{\CalP}$ can be deformed into the complex domain $L{\CalP}_{\BC}$, as a middle dimensional contour ${\Sigma}$, which splits, in the appropriate homology group, as a sum of Lefschetz thimble cycles associated
with the critical points of the exponential in \eqref{eq:trbeta}.  

 The classification of such cycles in one- and two-dimensional quantum field theories has recently been revived in e.g. \cite{Nekrasov:2018pqq,Krichever:2020tgp, Krichever:2021bdb}.

 String worldsheet metric is sometime Lorentzian, somewhere Euclidean, and likely complex in between  \cite{Witten:2013pra}.  Less often, integral, or quantized, parameters, such as the number $n$ of flavors or species (replica method), the number of colors $N$ (matrix models, master field), spacetime dimension $D$ (dimensional regularization), or angular momentum $l$ (Regge poles) are continued into the complex domain. For example, the character (twisted partition function) of a particle on $S^2$ (phase space) 
 \beq
 {\chi}_{j} ({\theta}) = \frac{{\rm sin}\left( (j+\frac 12) {\theta} \right)}{{\rm sin}(\frac{\theta}{2})}
 \eeq
 admits analytic continuation to complex $\theta$, $j$, etc. It was I.~Gelfand's dream to understand the dimensional regularization used in computations of Feynman diagrams in quantum field theory and the subsequent renormalization, through the representation theory of the non-compact Lie group $SL(2, {\BC})$ (whose real form is the Lorentz group of the four dimensional Minkowski space which looks like our space-time).

Theoretical physics often uses a trick of analytic continuation. In studying the disordered systems, where a free energy of a statistical mechanical model 
\beq
F(J) = - \frac{1}{\beta}{\rm log}Z({\beta}, J)
\eeq
is to be averaged, e.g. 
\beq
\langle F(J) \rangle_{J} = \int \, DJ\, e^{-\Vert J \Vert^2} \, F(J) 
\eeq
over some manifold of parameters $J$, e.g. in the spin glass models, where 
\beq
Z({\beta}, J) = \sum_{\{ {\sigma}_{i} = \pm 1 \} } e^{-{\beta} \sum\limits_{\langle i,j \rangle} J_{ij} {\sigma}_i {\sigma}_j}
\eeq
In replica trick, one computes the averages of $n$'th powers of $Z$, then takes the limit $n \to 0$ to extract the free energy. Another popular application of the replica trick is in the computation of von Neumann entropy 
${\rm Tr} {\rho} {\rm log} {\rho}$ which is extracted in the $n \to 1$ limit of ${\rm Tr} {\rho}^{n}$. Thus, one studies the statistical mechanics of $n$ copies of the original system. For integer positive $n$ this makes sense, but what is the meaning of the $n$ complex?

Curiously, we find several examples of systems for which these analytic continuations can be realized
 with conventional supersymmetric quantum field (or M-) theory.  The supersymmetry is softly broken near a defect supporting the theory being analytically continued in some of its parameters. We are interested in analytic continuations of 
 spins, dimensions, temperatures, fugacities etc.

 \section{Models with $O(N)$ symmetry}

 In our first model we study the conformally coupled scalar living on $S^{N-1}$.  Conformally coupled scalar field living on an $N$-dimensional 
Riemanian manifold $M^N$
is described by the action
\beq
S = \frac 12 \int_{M^N} d^{N}{\xi} \sqrt{g} \left( g^{mn} {\partial}_{m} {\varphi} {\partial}_{n} {\varphi} + \frac{N-2}{2N-2} R(g){\varphi}^{2} \right)
\eeq 
In this paper we are interested in the conformally flat manifolds of the form $S^{N-1} \times S^{1}$ (more general case $S^{N_{1}-1} \times S^{N_{2}-1}$ with $N_1, N_2 \geq 3$ is also connected to supersymmetric systems, but the story is
more complicated). We endow it with the metric
\beq
ds^2_{M^{N}} = {\beta}^{2} dt^{2} + R^{2} d{\Omega}_{S^{N-1}}^{2}
\eeq
with $t \sim t+ 1$,  
\beq
\frac{{\beta} R^{N-1} }{2} \int_{S^{N-1} \times S^{1}} dt d^{N-1} {\Omega} \, \left[ \frac{1}{{\beta}^{2}} \left( {\partial}_{t} {\varphi} \right)^2 + \frac{1}{R^{2}} \left( 
 {\nabla}{\varphi} \cdot {\nabla}{\varphi}  + \left( \frac N2  -1 \right)^{2} {\varphi}^{2} \right) \right]
\label{eq:confcs}
\eeq
We can decompose ${\varphi}(t,x) = \sum_{l=0}^{\infty} \sum_{A=1}^{d_{l}(N)} {\varphi}_{A,l}(t) Y_{A,l}(x)$ into the
spherical functions, where 
\beq
- {\Delta}_{S^{N-1}_{R}} Y_{A,l} = \frac{l(l+N-2)}{R^{2}} Y_{A,l} \ .
\eeq  
We see \eqref{eq:confcs} is a Lagrangian describing 
an infinite set $\left( {\varphi}_{A,l}(t) \right)_{l=0}^{\infty}{} _{A=1}^{d_{l}(N)}$ of  harmonic oscillators  with frequencies $\frac{l + \frac N2 - 1}{R}$. 
The multiplicity $d_{l}(N)$ is computed from the generating function 
\beq
\sum_{l=0}^{\infty} x^{l} d_{l}(N) = \frac{1+x}{(1-x)^{N-1}} 
\label{eq:genfnx}
\eeq
(this is $1-x^2$ times the character of the space of polynomials in $N$ variables, with $x$ being the fugacity for dilatation symmetry). 
Thus
\begin{multline}
d_{l}(N) =  
\left( \begin{matrix} N+l-2 \\ l \end{matrix} \right) + \left( \begin{matrix} N+l-3 \\ l-1 \end{matrix} \right) \ , \\
d_{l}(2) = 2 - {\delta}_{l,0}\, , \ d_{l}(3) = 2l+1\, , \ d_{l}(4) = (l+1)^2\, , \ldots  \label{eq:dln}
\end{multline}
Note that $d_{l}(N)$ admits analytic continuation to complex values of $N$, as
\beq
D_{r}(N) \, =\,  \frac{2 r}{{\Gamma}(N-1)} \frac{{\Gamma}(r+N/2-1)}{{\Gamma}(r-N/2+2)} \, , \qquad r = N/2-1+l
 \eeq
 from which the symmetry
 \beq
 D_{r}(N) = (-1)^{N} D_{-r}(N)
 \eeq
 is readily seen. This is related to the symmetry $x \mapsto x^{-1}$ of \eqref{eq:genfnx}
 accompanied by the $(-1)^{N} x^{N-2}$ prefactor. 
 
We can thus write the
partition function of conformal scalar on $S^{N-1}_{R} \times S^{1}_{\beta}$:
\beq
Z_{N}(q) = {\rm Tr}_{{\CalH}_{S^{N-1}_{R}}}  \, e^{-{\beta}{\hat H}} = q^{c_{0}} \prod_{r \geq \frac N2 - 1}^{\infty} \frac{1}{\left( 1 - q^{r}  \right)^{D_{r}(N)}} = q^{c_{0}} {\exp}\, \sum_{n=1}^{\infty} \frac{1}{n} \frac{q^{-n/2}+q^{n/2}}{(q^{-n/2}-q^{n/2})^{N-1}}
\label{eq:confspf}
\eeq
where $q$ is the conformal invariant:
\beq
q = e^{-\frac{\beta}{R}}
\eeq
and $c_{0}/R$ is the Casimir energy.\footnote{One can compute $c_0$ as a regularized sum
\beq
c_{0} = \frac 12 \sum_{l=0}^{\infty} d_{l}(N) \left( l + N/2-1 \right) :=   - \frac 12  {\rm Coeff}_{\tau^1}  \frac{e^{\tau/2}+e^{-\tau/2}}{(e^{\tau/2}-e^{-\tau/2})^{N-1}}\Biggr\vert_{\tau = 0}  
\eeq
See \cite{Cardy:1991kr} for more systematic treatment}
We can refine \eqref{eq:confspf} by introducing the general $O(N)$ twisted boundary conditions:
\beq
{\varphi}(t+1, x) = {\varphi}(t, {\bf g}\cdot x)
\eeq
where ${\bf g} \in O(N)$ is characterized by $[N/2]$ angles ${\theta}_{1}, \ldots, {\theta}_{[N/2]}$ and, for odd $N$,  in addition, by ${\xi} = \pm 1$:
\beq
\begin{cases} N = 2m \, , \qquad \qquad  Z_{N}(q, {\theta}_{1}, \ldots, {\theta}_{m}) = {\exp}\, \sum\limits_{n=1}^{\infty} \,  \, \frac{q^{-n}-q^{n}}{n} \prod\limits_{a=1}^{m} \frac{1}{q^{n}-2 {\rm cos}(n{\theta}_{a}) + q^{-n}}\\
N = 2m+1 \, , \qquad  Z_{N}(q, {\theta}_{1}, \ldots, {\theta}_{m}, {\xi}) = {\exp}\, \sum\limits_{n=1}^{\infty} \frac{
q^{-n/2}+q^{n/2}{\xi}^{n}}{n} \prod\limits_{a=1}^{m} \frac{1}{q^{n}-2 {\rm cos}(n{\theta}_{a}) + q^{-n}}\\
\end{cases}
\eeq
For three dimensional conformally coupled scalar this formalism gives:
\beq
Z(q, {\theta}, {\xi}) = {\exp}\, \sum\limits_{n=1}^{\infty} \frac{1}{n}  \frac{q^{-n/2}+q^{n/2}{\xi}^{n}}{q^{n}+q^{-n}-2 {\rm cos}(n{\theta})}
\eeq
One can get similar formulas for the conformally coupled fermions. 
These three-dimensional partition functions are closely related to MacMahon function
\beq
M(q) = \prod_{n=1}^{\infty} \frac{1}{(1-q^n)^n} \, = \, {\exp}\, \sum_{n=1}^{\infty} \frac{1}{n} \frac{q^{n}}{(1-q^{n})^2}
\label{eq:mcm}\eeq
which counts plane partitions \cite{Okounkov:2003}, is featured in representation theory of $W_{1+\infty}$-algebra \cite{Awata:1994tf, Awata:1994em}, and in the theory of topological strings \cite{Bershadsky:1993cx, Okounkov:2003sp}. 
Recall \cite{Nekrasov:2008} that the supersymmetric partition function of the $\Omega$-deformed maximally supersymmetric $U(1)$ Yang-Mills theory on noncommutative ${\BR}^{6}$ (or $D5-D(-1)$ system) is equal to
\beq
Z_{6}(q, {\ve}_{1}, {\ve}_{2} , {\ve}_{3}) = M(q)^{-\frac{({\ve}_{1}+{\ve}_{2})({\ve}_{1}+{\ve}_{3})({\ve}_{2}+{\ve}_{3})}{{\ve}_{1}{\ve}_{2}{\ve}_{3}}}
\label{eq:zmcm}
\eeq
where $q$ is the $D(-1)$-instanton fugacity, and ${\ve}_{1}, {\ve}_{2}, {\ve}_{3}$ are the three parameters of the $U(3)$ $\Omega$-deformation. The group $U(3) \subset O(6)$ rotates the space-time $\BR^6$ and the transverse $\BR^2$ via determinant $U(1) \approx O(2)$.    Thus, the replica of three dimensional bosons and fermions can be realized physically as six dimensional gauge theory by playing with the $\Omega$-deformation parameters. Once the $\Omega$-deformation preserves more supersymmetry, the exponent in \eqref{eq:zmcm} becomes $1$. 

Similarly, two dimensional boson partition function ($N=2$) 
\beq
Z(q) = q^{-\frac{1}{12}} \prod_{n=1}^{\infty} \frac{1}{(1-q^n)^2} 
\eeq 
is related to Euler function
\beq
{\phi}(q) = \prod_{n=1}^{\infty} (1-q^{n})
\eeq
The replica of the two dimensional chiral bosons is given by the $\Omega$-deformed ${\CalN}=2^*$ four dimensional $U(1)$ super-Yang-Mills theory: 
\beq
Z_{4}(q, {\ve}_{1}, {\ve}_{2} , {\ve}_{3}) = \left( \frac{1}{{\phi}(q)} \right)^{\frac{({\ve}_{1}+{\ve}_{3})({\ve}_{1}+{\ve}_{2})}{{\ve}_{1}{\ve}_{2}}}
\eeq
where $\ve_3$ is the mass of the adjoint hypermultiplet \cite{Nekrasov:2003rj, Nekrasov:2015wsu}. 
One can refine the ``replica theories'' by lifting them to the supersymmetric theories in one dimension higher, compactified on a circle. The $\ve$-parameters become $q$-parameters. For example, the two dimensional example refines to
\beq
Z_{5}(q, q_1, q_2, q_3) \, = \, {\exp}\, \sum_{n=1}^{\infty} \frac{q^{n}}{n} \frac{(1-q_{1}^{n} q_{3}^{n})(1-q_{2}^{n}q_{3}^{n})}{(1-q_{1}^{n})(1-q_2^{n})(1-q^{n})}
\eeq
which has two interpretations: as Witten index of five dimensional maximally super-symmetric Yang-Mills theory
compactified on a circle with the Lorentz and R-symmetry twists compatible with a fraction of supersymmetry, or as elliptic genus of a six dimensional $(2,0)$-superconformal tensor theory compactified on a torus \cite{Nekrasov:2008}. 
The refinement of the replica of three dimensional theory brings about the (linearized) eleven dimensional supegravity on ${\BR}^{10} \times S^1$, subject to an $SU(5)$ twist \cite{Nekrasov:2008, Nekrasov:2014nea}.  Recently this partition function was reproduced using a twisted version of supergravity \cite{Raghavendran:2021qbh}. 

Our second problem related to $S^{N-1}$ is the classical two dimensional $O(N)$-model, i.e. the sigma model with $S^{N-1}$ as a target. We found, in \cite{Krichever:2020tgp,Krichever:2021bdb}, that the complex critical points of the 
action 
\beq
\int_{T^2} \, \sqrt{h} h^{ab} {\nabla}_{a} {\vec n} \cdot {\nabla}_{b} {\vec n}
\eeq
with ${\vec n} \in {\BC}^{N}$, ${\vec n} \cdot {\vec n} = \sum_{i=1}^{N} n_{i}^{2} = 1$, and complex flat metric $h_{ab}$ on $T^2$ are described in terms of an algebraic integrable system.  The vector ${\vec n}$ with $N$ components is recovered from the so-called Baker-Akhiezer $\psi (k)$ function, which is an analytic function on an algebraic curve (called the Fermi-curve). The components $n_i$ are related to the evaluation of $\psi$ at $N$ special points, singled out by an additional structure. In this way all $O(N)$ models are unified, and in a sense $N$ is complexified (replaced by an algebraic curve).

\section{Interpolating between $SU$, $SO$, $Sp$}

In Dyson's original work on ensembles of random matrices with various time-reversal properties
\cite{Dyson:1962} one finds a discrete series of measures on the space of $N$ indistinguishable 
particles on a real line (or a circle), which  is naturally generalized into a continuous family of
$\beta$-ensembles. The points $\beta = 1,2,4$ correspond to the unitary, orthogonal, and symplectic groups. The interpolating model is no longer a matrix model, yet it can be viewed, rather artificially, as a double-trace deformation of a more conventional matrix model. 

In a simplest setting, the partition function of the $\beta$-ensemble is given by the integral
\beq
Z_{\beta} = \int_{{\BR}^{N}/S(N)} dx_{1} \ldots dx_{N} \, \prod_{i < j} | x_{i} - x_{j} |^{2\beta} 
\, e^{-\sum_{i=1}^{N} V(x_{i})}
\label{eq:betens}
\eeq
with some single-particle potential $V$. The beauty of \eqref{eq:betens} is that these models share many features with the fractional quantum Hall systems, they are amenable to the analysis  via
powerful Ward identities, more recently they were connected
\cite{Sulkowski:2009ne, Chekhov:2010zg, Nishinaka:2011aa}
 to instanton partition functions \cite{N2} and topological recursion \cite{Eynard:2008we}. The connection between the partition sums
of supersymmetric gauge theory and random matrix models was pointed out in \cite{Nekrasov:2003rj} for ${\ve}_{1} = - {\ve}_{2}$ background, where it was suggested that the ${\CalN}=1$ Dijkgraaf-Vafa theories \cite{Dijkgraaf:2002dh} solved by matrix models could potentially be analyzed systematically starting with the ${\CalN}=2$ instanton count. It is however puzzling that the matrix models, at least in the asymptotic large $\hat N$ expansion, admit the $\beta$-deformation, while the $\CalN=1$ theory seems to be confined to the anti-self-dual $\Omega$-backgrounds. In a parallel series of developments, \cite{Dijkgraaf:2009pc} proposed to describe the \emph{refined topological strings} using $\beta$-deformed matrix models. 

All these examples above get embedded in a conventional supersymmetric gauge theory setup (and string/M-theory via \cite{Nekrasov:2014nea}) via identification
\beq
{\beta} = - \frac{\ve_1}{\ve_2}\, , {\rm or} \ - \frac{\ve_2}{\ve_1}
\eeq
The simplest way to understand the emergence of matrix-model like structure is by looking at the instanton measure written in the plethystic form \cite{Nekrasov:2015wsu}, and observe that
the contribution
\beq
I_{\rm vm} = {\sf E} \left[  - \frac{S_{12} S_{12}^{*}}{(1-q_{1}^{-1}) (1-q_{2}^{-1})} \right]
\label{eq:vmcontr}
\eeq
of the vector multiplets can be written as:
\beq
I_{\rm vm} = {\sf E} \left[  - \frac{(1 - q_{1} ) M_{12} M_{12}^{*}}{(1-q_{2}^{-1})} \right]
\label{eq:vmmeasure}
\eeq
where we write
\beq
S_{12} = N_{12} - K_{12} (1-q_1)(1-q_2) = M_{12} (1-q_{1})
\eeq
(see \cite{Nekrasov:2015wsu} for explanations of the notations) where, importantly, $M$ is the pure character:
\beq
M_{12} = \sum_{\alpha=1}^{n} \sum_{j=1}^{\infty} e^{a_{\alpha}} q_{2}^{j-1} q_{1}^{{\lambda}^{({\alpha})t}_{j}} \sim \sum_{I} e^{x_{I}}
\eeq
where $x_{I}$ will play the role of the eigenvalues of some \emph{effective matrix} ${\CalX}$. In the limit where all $\ve$'s, $a$'s
are therefore $x_I$'s are uniformly scaled to zero, the measure \eqref{eq:vmmeasure}
approaches the $\beta = - {\ve}_{1}/{\ve}_{2}$-deformed matrix ensemble for $\CalX$:
\beq
{\Delta}_{\CalX}^{2\beta} :=  \prod_{I \neq J} ( x_{I} - x_{J} )^{- {\ve}_{1}/{\ve}_{2}}
\eeq
where we are being cavalier with the infinite-dimensional products. 
Similar considerations apply to the contributions of the bi-fundamental and fundamental hypermultiplets. Thus quiver gauge theories become multi-matrix models. 

One application of this imprecise dictionary is the matrix interpretation of the folded and crossed instantons of \cite{Nekrasov:2015wsu} and followup papers. For example, the folded instanton
setup fuses ${\CalN}=2^{*}$ gauge theory on ${\BR}^{4} = {\BC}^{2}$ with $\Omega$-background with parameters ${\ve}_{1}, {\ve}_{2}$, and mass of the adjoint hypermultiplet ${\ve}_{3}$ with the ${\CalN}=2^{*}$ theory on ${\BC}^{2}$ with $\Omega$-background with parameters ${\ve}_{2}, {\ve}_{3}$ and adjoint mass ${\ve}_{1}$. A typical instanton measure in this case looks like
\beq
{\sf E} \left[  - (1-q_{3}) \frac{S_{12} S_{12}^{*}}{(1-q_{1}^{-1}) (1-q_{2}^{-1})} - (1-q_{1}) \frac{S_{23} S_{23}^{*}}{(1-q_{2}^{-1}) (1-q_{3}^{-1})}  + q_{3} \frac{1-q_{1}^{-1}q_{2}^{-1}q_{3}^{-1}}{1-q_{2}^{-1}} S_{12}S_{23}^{*} \right]
\label{eq:folded}
\eeq
Now, introducing the effective matrices ${\CalX}, {\CalY}$ whose eigenvalues $(x_I)$, $(y_J)$
are defined through
\beq
\sum_{I} e^{x_{I}}  = \frac{S_{12}}{1-q_{1}} \, , \ \sum_{J} = \frac{S_{23}}{1-q_{3}}
\eeq
we rewrite \eqref{eq:folded} as
\beq
\frac{{\Delta}_{\CalX}^{2\beta} {\Delta}_{\CalY}^{2/{\beta}}}{{\rm Det}(ad({\CalX}) + {\ve}_{3})^{\beta} \, {\rm Det}(ad({\CalY}) + {\ve}_{3})^{1/{\beta}}} \times \frac{{\rm Det}({\CalX} \otimes 1 - 1 \otimes {\CalY} + {\ve}_{3})
{\rm Det}({\CalX} \otimes 1 - 1 \otimes {\CalY} - {\ve}_{3})}{{\rm Det}({\CalX} \otimes 1 - 1 \otimes {\CalY})^{2}}
\eeq
in other words, a $\beta$-deformed supermatrix $3$-matrix model (cf. \cite{KKN})  with the superpotential
\beq
{\rm Tr} \left(  {\CalZ} [ {\CalX}, {\CalY} ] + {\ve}_{3} {\CalZ}^{2} + V({\CalX}) + W ( {\CalY}) \right)
\eeq  
\section{Complex spins}

Spin chains, originally introduced  to model the magnetic phenomena in metals, later generalized to models of ice etc. typically  involve finite-dimensional representations of the spin group, e.g. $SU(2)$. In recent years an interest to spin chains with infinite dimensional representations has been growing. One source of interest is the integrable structure of two-dimensional conformal field theory \cite{Bazhanov:1994ft}, quantum hydrodynamics \cite{AO:2013,Litvinov:2013zda}.
Another source of interest is the integrable structure discovered in the planar ${\CalN}=4$ super-Yang-Mills in four dimensions, where a spin chain based on the non-compact supergroup $PSU(2,2|4)$ is detected \cite{Beisert:2010jr}. 
Finally, a growing body of evidence is provided by the supersymmetric sector of the $\Omega$-deformed ${\CalN}=2$
theories in four dimensions (and their five and six dimensional lifts). Namely, a mass deformed superconformal quiver theory in four dimensions is expected to be related to an integrable spin chain based on the Yangian of the Kac-Moody algebra associated with the quiver. Instead of stating the general conjecture, let us give a specific example: 
a super-QCD ${\CalN}=2$ theory with $SU(N)$ gauge group, with $2N$ hypermultiplets in fundamental representation, 
with $\Omega$-deformation in two out of four dimensions, is related to the $\mathfrak{sl}_{2}$ spin chain, with the
space of states being a subspace of the tensor product of $N$ infinite-dimensional modules :
\beq
H = V_{s_{1}, a_{1}} (u_1) \otimes V_{s_{2}, a_{2}} (u_2) \otimes \ldots \otimes V_{s_{N}, a_{N}} (u_N)
\label{eq:hvv}
\eeq 
Here $V_{s,a}$ denotes a representation of the algebra $\mathfrak{sl}_{2}$, which has a eigenbasis $|n\rangle$ labelled by $n\in \BZ$, with
\beq
L_{0} |n \rangle = (n +a) | n \rangle\, , \ L_{+} | n \rangle = (n + a - s ) | n+1 \rangle\, , \ L_{-} |n \rangle = ( n+a+s ) |n-1 \rangle
\eeq
The parameter $s \in \BC$ is the spin of the representation, related to the Casimir by $L_0^2 - \frac 12 \left( L_{+}L_{-} +  L_{-}L_{+} \right) = s(s+1)$. The parameter $a \in \BC$ determines a mid-point in the spectrum of $L_0$. The familiar finite-dimensional representations of $SU(2)$ are found for half-integer $s$ and $a -s$, $a+s$ both non-negative integers.  
 
The gauge theory corresponds to $A_1$ quiver. The $a_i$'s are the Coulomb moduli, i.e. the eigenvalues of the adjoint scalar in the vector multiplet, while the spins $s_i$'s are related to the masses by $m_{i}^{+} - m_{i}^{-} = 2 s_i$, $i = 1, \ldots , N$. The sums $u_{i} = \frac 12 \left( m_{i}^{+} + m_{i}^{-} \right)$ are the inhomogeneities, the evaluation parameters of the Yangian modules based on the $\mathfrak{sl}_2$ modules.   Although the permutations of the $2N$ masses is a symmetry of the four dimensional theory, the specific two dimensional subsector defined by $\Omega$-deforming and adding a surface defect, depends on the ordering of masses, which, in turn, defines the spins etc. See \cite{Lee:2020hfu,Jeong:2021rll,Jeong:2023qdr} for more details. 

Another view on the same surface defect in the fully $\Omega$-deformed theory reveals the connection to Knizhnik-Zamolodchikov equation borrowed from the $\widehat{\mathfrak{sl}_{N}}$-representation theory \cite{Nekrasov:2021tik,Jeong:2021rll,Jeong:2023qdr}. Again, complex spins and infinite-dimensional modules feature prominently.

\section{Future directions}

It would be nice to turn on interactions in the two and three dimensional theories. The supersymmetric methods of a different kind (Wegner and Efetov) are used in the study of disordered systems for many decades \cite{MZ}. It would be nice to find a relation between our higher dimensional theories with conventional supersymmetry and the supercoset sigma models used in the studies of integer Hall plateau transitions \cite{MZ2}. 

The supersymmetric interfaces \cite{DN} in one, two, and three dimensional supersymmetric gauge theories with eight supercharges realize finite dimensional spin chains. A $T$-dual construction
realizes the same systems in a four dimensional version of Chern-Simons theory \cite{Costello:2013sla}. It would be useful to find a four dimensional analogue of the interfaces \cite{DN} providing a geometric realization of $R$-matrix suitable for infinite dimensional representations of quiver Yangians and quantum loop algebras. 

The most interesting and mysterious question is the role of complex metrics in string theory. In addition to the arguments of \cite{Witten:2013pra} one would like to use the complex worldsheet metric to avoid unphysical poles in the bulk of ${\CalM}_{g}$ which pop up in time-dependent cosmological string backgrounds  \cite{Nekrasov:2002kf}, as well as the spacetime considerations of \cite{Halliwell:1989dy, Kontsevich:2021dmb,Witten:2021nzp}. We hope to return to this question in the future (whatever it means in the complex world).

\section{Acknowledgements}

Discussions with M.~Dedushenko, A.~Grekov, S.~Jeong, N.~Lee, A.~Okounkov, N.~Seiberg, O.~Tsymbaliuk, are gratefully acknowledged. Many thanks to the organizers of String-Math'2002 in Warsaw, especially to Piotr Sulkowski, for providing a stimulating working environment in troubled times. Research was partly supported by the National Science Foundation by award NSF PHY Award 2310279. Any opinions expressed are solely our own and do not represent the views of the National Science Foundation.

%%%%%%%%%%%%%%%%%%%%%%%%%%%%%%%%%%%%%%%%%%%%%%%%%%%%%%%
%%%%%%%%%%%%%%%%%%%%%%%%%%%%%%%%%%%%%%%%%%%%%%%%%%%%%%%

\end{document}